\newcommand{\half}{\frac{1}{2}}
\newcommand{\adag}{a^{\dag}}

 \documentstyle[preprint,aps,tighten]{revtex}
\begin{document}
\draft
\title{Interpolating statistics realized as Deformed Harmonic Oscillators}
\author{ P. Narayana Swamy }
\address{Department of Physics, Southern Illinois University,
Edwardsville IL 62026}

\maketitle
\begin{abstract}
The idea that a system obeying interpolating statistics can be
described by a deformed oscillator algebra has been an outstanding
issue. This original concept introduced long ago by Greenberg is
the motivation for this investigation. We establish that a
q-deformed algebra can be used to describe the statistics of
particles (anyons) interpolating continuously between Bose and
Fermi statistics, i.e., fractional statistics. We show that the
generalized intermediate statistics splits into the Boson-like and
Fermion-like regimes, each described by a unique oscillator
algebra. The B-anyon thermostatistics is described by employing
the q-calculus based on the Jackson derivative but the F-anyons
are described by ordinary derivatives of thermodynamics.
Thermodynamic functions of both B-anyons and F-anyons are
determined and examined.
\end{abstract}

 \vspace{2.8in}
Electronic address: pswamy@siue.edu \vspace{.2in}
 \indent December 2004


\section{Introduction}

The statistical mechanics of anyons existing in 2+1 space-time
dimensions was investigated in detail sometime ago\cite{RAPNS1},
where many of the thermodynamic properties of these particles,
such as the partition function, entropy, pressure, internal
energy, virial expansion and specific heat were determined. The
analysis of thermostatistics in that investigation was performed
in two space dimensions based on a simple ansatz for the anyon
distribution function. In particular it was found that a natural
distinction arises between boson-like and fermion-like anyons.
Virial coefficients of the two dimensional anyon gas were
determined and compared with earlier investigations on the subject
\cite{earlier}. More recently, the subject of generalized
statistics has been studied in other investigations
\cite{Frappat,Frau}, where oscillator algebras and field theory in
low space dimensions are employed. There has also been great
interest in a comparative study of different theories of
interpolation \cite{Chatur} in the context of Haldane and Gentile
statistics .

\vspace{.2in}

Physical systems in 2+1 space-time dimensions display many unusual
quantum properties, especially relating to rotation and spin.
These features lead to interesting results for the quantum
mechanics of angular momentum in two space dimensions  giving rise
to fractional spin, and hence not quantized in familiar half
integer multiples of the Planck constant. These objects are named
in the literature as anyons, representing fractional statistics
which accordingly interpolate between Bosons and Fermions
determined by the relation defining permutation symmetry for the
many body wave function
\begin{equation}\label{1}
    \psi(x_1, \cdots x_j \cdots x_i \cdots , x_n) = e^{i \pi \alpha }
\psi(x_1, \cdots x_i\cdots x_j \cdots , x_n)\, ,
\end{equation}
where the statistics determining parameter $\alpha$ in this
relation is a real number and the limits $\alpha = 0,1$ correspond
respectively to Bosons and Fermions. Since the permutation
symmetry is related to rotations in two space dimensions, anyons
may be described by the braid group, described by the infinite
group of $N$ strands represented by products of Pauli matrices
\cite{Lerda}. This connection between fractional spin and
 fractional statistics, with Chern-Simons Field theory providing a
representation for the anyons,  has been the reason why anyons
have been thought to exist only in 2+1 dimensions. It must,
however, be stressed that permutation symmetry or the exchange
symmetry is more general than what arises from rotations.

\vspace{.2in}

Indeed, a theory of interpolating or intermediate statistics has
recently been formulated which provides a definite clarification
of this issue. In this investigation \cite{RAPNS2}, a theory of
the exchange symmetry of many particles is developed so as to
allow a continuous interpolation between Bose and Fermi
statistics. This theory has many interesting features, one of them
being that the system is not restricted to 2+1 dimensions. The
other features are: interpolating statistics arises directly from
principles of thermodynamics, specifically the principle of
detailed balance; basic numbers arise naturally and automatically
in this theory; the theory reproduces standard expressions in the
Bose and Fermi limits; the general expression for the distribution
function is the solution of a transcendental equation which can be
expressed as a power series and also in the form of a continued
fraction; the first approximant of this solution agrees with an
approximate distribution function introduced as an ansatz in the
thermostatistics of anyons investigated earlier\cite{RAPNS1}.

\vspace{.2in}

There is another generalization of the standard thermostatistics
that has been studied extensively  in the literature which has to
do with the theory of $q$-deformed quantum oscillators i.e.,
quantum groups \cite{Bieden}. These are objects described by a
deformed algebra of the harmonic oscillators signified by a
parameter $q$ so that $q=1$ corresponds to the boson oscillators.
It has been shown that the $q$-calculus based on the Jackson
derivative (JD) can be employed rather than the ordinary partial
derivatives in order to describe the thermostatistics of
$q$-bosons\cite{ALPNS1}. Indeed it has been shown that the
complete theory of generalized thermodynamics of $q$-bosons and
$q$-fermions can be developed using basic numbers with the base
$q$ and employing the $q$-calculus based on JD. The thermodynamic
functions such as entropy, pressure, internal energy, specific
heat etc. of such systems have been studied\cite{ALPNS2} and
compared with ordinary bosons and fermions.

\vspace{.2in}

As the generalized thermodynamics of generalized bosons and
fermions can  be successfully described by a theory based on the
$q$-deformed algebra, we can accordingly pose the following
question: What kind of deformed algebra would describe
interpolating statistics, the statistics of anyons? This question
is even more interesting due to the fact that the recent
derivation of interpolating statistics from the principle of
detailed balance \cite{RAPNS2} introduces the basic numbers
naturally but without the benefit of an oscillator algebra which
leads to these basic numbers. Attempts have been made by Greenberg
and others\cite{Greenberg} to investigate $q$-deformed algebras as
possible foundations of interpolating statistics.

\vspace{.2in}

At one time it was regarded as conventional wisdom that
generalized or interpolating statistics can have no relation to
the algebra of deformed harmonic oscillators since oscillators
exist in any dimensions \cite{Lerda}. The connection with quantum
groups was originally discovered in \cite{LerdaSciuto} and
subsequent investigations have employed quantum groups in lower
dimensions\cite{Frappat,Frau}. It has been pointed out that the
issues of quantum statistics are very different from commutation
and anticommutation relations for local fields\cite{Lerda}. It has
been conjectured that quantum groups might be the characteristic
symmetry structures of anyon systems even though explicit
realizations of this fact are still missing and thus a direct
relation between the two should exist \cite{Frau}. If anyons are
described in two dimensions, and since the real world is of three
space dimensions, anyons may not be real particles. All of this
might be an indication that there ought to be an approach to
interpolating statistics other than from spin and rotation and
this is primarily the goal of the present investigation.

\vspace{.2in}

Anyons  might well be composites from charged particles and
magnetic flux tubes. In what follows we shall use the term
\emph{anyons},
 to refer to particles
 obeying  interpolating statistics. We shall use this term
  in its generic sense and not as particles described by any
   particular  model
  investigated in the literature and not restricted to
   2+1 space dimensions. This has indeed been an
   outstanding problem and never resolved until the present investigation.
     The fundamental question that
  comes to mind is:
Can a deformed oscillator algebra describe interpolating
statistics? If so, what kind of algebra would be most appropriate
to describe a statistical theory which would interpolate between
Bose and Fermi statistics? It is indeed the spirit of current
wisdom that interpolation might be considered in some sense as a
deformation and thus we expect the two aspects to be closely
related. The investigation of several possible algebras leads one
to conclude that many of these are not satisfactory: some do not
have the correct Bose or Fermi limits; some are inconsistent;
others do not lead to suitable real distribution functions etc.
Recently a successful formulation of the theory of $q$-deformed
bosons and fermions  was based on the premise that while
$q$-bosons and $q$-fermions arise from different oscillator
algebras, they can be considered as special cases of one
fundamental algebra. We shall accordingly begin with this algebra
which may be defined by
\begin{equation}\label{2}
    a \adag - \kappa q^{\kappa} \adag a = q^{-N}, \quad 0 \leq q \leq 1\, ,
\end{equation}
where $\kappa = 1$ will describe boson-like anyons, and $\kappa =
-1$ will be used to describe fermion-like anyons. Here $N$ is the
number operator which acts on the Fock space of oscillator states.
We shall introduce the abbreviations B-anyons and F-anyons for
simplicity. This algebra was investigated for arbitrary parameter
$q$ in \cite{ALPNS2} and $\kappa$ was taken to be either 1 or -1
to describe $q$-bosons and $q$-fermions. In the present work, we
shall investigate the B-anyons and F-anyons, corresponding to
$\kappa=1$ and $ \kappa=-1$ separately arising from the
corresponding algebra and treat $q$ as a parameter which will
provide the interpolation between the Bose and Fermi cases. It
must be stressed that the interpretation in the present work is
different and the consequences are also quite different
from\cite{ALPNS2}.

\section{The algebra of B-anyons}

We shall examine the algebra defined  by
\begin{equation}\label{3}
a \adag -  q \adag a = q^{-N}, \quad 0 \leq q \leq 1\, ,
\end{equation}
where $N$ is the number operator and $q\rightarrow 1$ corresponds
to the Bose limit. We may also add the relations
\begin{equation}\label{4}
    [N,a]=-a, \quad [N, \adag]= \adag\, ,
\end{equation}
which can be established by examining the action on Fock states.
It is well-known that this algebra leads to the introduction of
basic numbers. Nevertheless, it is instructive to briefly review
the analysis.

\vspace{.2in}

Let us define the operator $\tilde{N} = \adag a$ and its action on
the Fock state by $\tilde{N} |n\rangle = \alpha_n |n \rangle $,
assuming that the eigenvalue depends on $n$. First we derive the
algebra $\tilde{N} \adag - q \adag \tilde{N} = \adag q^{-N}$ as a
consequence of the algebra in Eq.(\ref{3}). This leads to the
recurrence relation $\alpha_{n+1}= q^{-n} + q \alpha_n$, the
solution of which  is immediately seen to be
\begin{equation}\label{5}
\alpha_n = q^{n-1} + q^{n-3} + \cdots + q^{-n +3}+ q^{-n+1}\, .
\end{equation}
This is recognized as the familiar basic number
\begin{equation}\label{6}
    \alpha_n = [n]=\frac{q^n - q^{-n}}{q - q^{-1}}\, ,
\end{equation}
with the corresponding operator form defined by
\begin{equation}\label{7}
    [N]= \frac{q^N - q^{-N}}{q - q^{-1}}\,.
\end{equation}
In this manner we see that the basic number $\adag a = [N]$, which
is different from the number operator,  arises  as a consequence
of the algebra in Eq.(\ref{3}) and is uniquely related to this
algebra.

\vspace{.2in}

In order to build the Fock space, we first observe the results $a
|n \rangle = \sqrt{\alpha_n}\, |n-1 \rangle$ and $\adag |n \rangle
= \sqrt{\alpha_{n+1}}\, |n+1 \rangle$.  We may then proceed to
construct the Fock states in the familiar straightforward manner,
thus obtaining
\begin{equation}\label{8}
    |n \rangle = \frac{(\adag )^n} {[n]!} |0 \rangle \, ,
\end{equation}
where
\begin{equation}\label{9}
    [n]!= [n] [n-1] \cdots [2] \cdot 1 \, .
\end{equation}
All of these reduce to the standard results of boson oscillators
in the Bose limit. We note that there are no restrictions on the
Fock states, with $n$ corresponding to any integer.

\vspace{.2in}

To proceed further, we introduce the Hamiltonian
\begin{equation}\label{10}
H = \sum_i N_i (E_i - \mu)\, .
\end{equation}
We note that contrary to appearances, this Hamiltonian does indeed
incorporate interpolation (or deformation), since the occupation
number depends on the parameter $q$ as we shall see. From the
definition of the mean value
\begin{equation}\label{11}
    q^{\pm n_k} = \frac{1}{\cal Z} {\rm Tr} (e^{-\beta H}q^{\pm
    N_k})\, ,
\end{equation}
we derive
\begin{equation}\label{12}
    [n_k]= \frac{1}{\cal Z}{\rm Tr} (e^{-\beta H}
    [N_k])\, .
\end{equation}
From the cyclic property of the trace and the property $f(N) \adag
= \adag f(N+1)$, valid for polynomial functions, we obtain the
result
\begin{equation}\label{13}
    e^{\beta (E_k-\mu)}= \frac{q^{-n_k}+ q [n_k]} {[n_k]}\, ,
\end{equation}
where the right hand side can be expressed in terms of $[n_k +1]$.
This leads to the solution for the mean occupation number:
\begin{equation}\label{14}
    n_i = \frac{1}{\ln q}
    \ln \left (
    \frac{e^{\beta (E_i - \mu)} - q^{-1} }
    {e^{\beta (E_i - \mu)} - q}
     \right )\, .
\end{equation}
This is a familiar result for q-boson oscillators, signifying the
effects of  $q$-deformation\cite{ALPNS2}. Here, however, we are
interpreting $q$ as the parameter   of interpolating statistics.
This is seen to reduce to the ordinary Boson distribution function
in the Bose limit.

\vspace{.2in}

The theory of $q$-deformed bosons is also characterized by the
Bargman-Wigner holomorphic representation of the annihilation and
creation operators, namely
\begin{equation}\label{15}
a \Longleftrightarrow {\cal D}^{q}(x)\, , \quad x
\Longleftrightarrow \adag \, ,
\end{equation}
where ${\cal D}^{q}(x)$ denotes the Jackson derivative (JD)
defined by
\begin{equation}\label{16}
{\cal D}^{q}(x) f(x) =  \frac{f(qx) - f(q^{-1}x)}{x(q-q^{-1})}\, .
\end{equation}
The JD arises naturally in q-deformed algebras \cite{Exton} and is
intimately linked with basic numbers. It has also been shown that
the connection of the basic number with JD may be attributed to
deformed Heisenberg algebra \cite{PNS2}. It has been further
established\cite{ALPNS2} that the structure of thermostatistics is
preserved if ordinary derivatives of thermodynamics are replaced
by JD\cite{ALPNS1,ALPNS2}. Accordingly, the various thermodynamic
quantities of interest can be obtained by employing the JD. The
results obtained in\cite{ALPNS2} are therefore readily seen to
apply to the interpolating statistics for the case of B-anyons.
Specifically we have to set $\kappa = 1$ in the results of
\cite{ALPNS2}. We shall now summarize these results for the
various thermodynamic functions.

\vspace{.2in}

The logarithm of the partition function is given by
\begin{equation}\label{17}
    \ln {\cal Z}= - \sum_i \ln (1- z e^{-\beta E_i})\,
\end{equation}
where $z = e^{\beta \mu}$ is the fugacity, as inferred from
\begin{equation}\label{18}
    N \; = \; \sum_i n_i = \; z {\cal D}^{(q)}(z) \ln {\cal Z}\, .
\end{equation}
The $q$-dependence arises after employing the JD.  To obtain  the
various thermodynamic functions, it is convenient to replace the
sums over states by integrals in the usual manner
\begin{equation}\label{19}
    \sum_i \Longrightarrow V (2 \pi)^{-3}\int \; d^3 k \, ,
\end{equation}
where $V$ is the volume, and introduce\cite{Reichl} the thermal
wavelength $\lambda = h / \sqrt{2 \pi m k T}$. In this manner, we
obtain the expression for the internal energy
\begin{equation}\label{20}
    U = \frac{3}{2 \lambda^3}\,V T g_{5/2}(q,z)\, ,
\end{equation}
where
\begin{equation}\label{21}
    g_{n}(q,z)= \frac{1}{q-q^{-1}}\, \left ( \sum_{r=1}^{\infty}
     \frac{(q z)^r} {r^{n+1}} -
     \sum_{r=1}^{\infty}
     \frac{(q^{-1} z)^r} {r^{n+1}}\right ) \, ,
\end{equation}
corresponds to the familiar \cite{RAPNS1,ALPNS2} generalized
Riemann Zeta functions involving both $q$ and $q^{-1}$, which
reduces to the standard $g_n$ function in the Bose limit. The
entropy of the B-anyons is given by
\begin{equation}\label{22}
    \frac{S}{V}= \frac{1}{\lambda^3} \left \{
    \frac{5}{2}\, g_{5/2}(q,z) - g_{3/2}(q,z) \ln z
    \right \}\, .
\end{equation}
Recalling the range $0 \leq q \leq 1$, we observe that the
thermodynamic quantities such as the internal energy, entropy etc.
for the B-anyons are smaller than for ordinary Bosons. This
follows from the known behavior of these functions\cite{Reichl} as
a function of the argument. The pressure of the B-anyon gas is
given by
\begin{equation}\label{23}
    \frac{P}{T}= \frac{1}{\lambda^3}\, g_{3/2}(q,z)\, ,
\end{equation}
which is discussed in\cite{ALPNS2}.

\vspace{.2in}

Finally, considering that the effects due to the statistics
parameter $q$ are rather subtle, we would like to consider, as an
illustration,  the virial expansion in the case of B-anyons in
some detail. We begin with the result
\begin{equation}\label{24}
    \frac{\lambda^3}{v}= g_{3/2}(z,q)\, ,
\end{equation}
where $v= V/N$. The generalized function here can be conveniently
expressed in the form of the series
\begin{equation}\label{25}
g_{3/2}(z,q)= z + \frac{[2]}{2^{5/2}}\, z^2 +
\frac{[3]}{3^{5/2}}\, z^3 + \cdots \, ,
\end{equation}
and it is easy to see that it reduces to the ordinary function
$g_{3/2}(z)$ in the limit $q \rightarrow 1$. Now for $q\neq 1$, we
have the result in Eq.(\ref{24}) given  by the above series, which
can be reverted to obtain
\begin{equation}\label{26}
z = (\frac{\lambda^3}{v}) - \frac{[2]}{2^{5/2}}\, \left (
\frac{\lambda^3}{v} \right ) ^2 + \left ( \frac{[2]^2}{2^2}-
\frac{[3]}{3^{5/2}} \right )\, \left ( \frac{\lambda^3}{v} \right
) ^3 + \cdots \, .
\end{equation}
Since
\begin{equation}\label{27}
    \frac{Pv}{kT}= \frac{v}{\lambda^3}\, g_{5/2}(q,z)\, ,
\end{equation}
upon substitution, we immediately obtain the virial expansion
given by
\begin{equation}\label{28}
\frac{Pv}{kT} = 1 - \frac{[2]}{2^{7/2}} \, \frac{\lambda^3}{v} +
\cdots \, .
\end{equation}
In the Bose limit, this reduces to the standard virial expansion.
However, the above result shows that for B-anyons, for $q\neq 1$,
the virial coefficients  generally involve the basic numbers and
hence $q$ dependent. On the basis of the second term, we might
state that the second and other virial coefficients for the
B-anyons are $q$-dependent and larger than for standard Bosons.

\vspace{.2in}

 In conclusion, the algebra in Eq.(\ref{3}) describes the
interpolating statistics of Boson-like particles, B-anyons. The
system is described in terms of JD and the basic numbers. All the
thermodynamic functions can be determined in terms of the
statistics determining parameter $q$ and the properties of the
gas, such as the equation of state etc.,  can be described as in
\cite{ALPNS2}.

\section{B-anyon distribution function}

We may further investigate the distribution function of the
B-anyons  as follows. We notice that the logarithmic form in
Eq.(\ref{14}) enables us to express the distribution function in
the form
\begin{equation}\label{29}
    n_i= - \frac{1}{2 \ln q^{-1}}\; \ln \left( 1-
    \frac{q^{-1}-q}{e^{\eta_i}-q} \right ) \, ,
\end{equation}
where $\eta_i = \beta (E_i - \mu)$. We may then express this in
the form of a continued fraction,
\begin{equation}\label{30}
    n_i=
    \left(  \frac{1}{2 \ln q^{-1}} \right )\; {
     y_i\over \displaystyle 1 -
     {\strut {1}^2 \, y_i \over\displaystyle 2-
     {\strut {1}^2\,  y_i
     \over\displaystyle 3 -
{\strut {2}^2 \, y_i \over\displaystyle 4 -
      \cdots }}}}\, ,
\end{equation}
where $y_i= (q^{-1}-q)/ (e^{\beta(E_i - \mu)} -q )$. We note that
$q^{-1} > q$. The infinite continued fraction form \cite{CF},
besides being a convenient tool to express an infinite series in a
compact form, has many desirable properties. For instance we
observe that the first two convergents (approximants) are given by
\begin{equation}\label{31}
    n_i^{(1)}= \left ( \frac{q^{-1}-q}{2 \ln q^{-1}}\right ) \;
    \frac{1}{e^{\displaystyle \eta_i}-q}\, ,
\end{equation}
and
\begin{equation}\label{32}
 n_i^{(2)}= \left ( \frac{q^{-1}-q}{2 \ln q^{-1}} \right ) \;
    \frac{1}{e^{\displaystyle \eta_i}-(\displaystyle \frac{q + q^{-1}}{2})}\, .
\end{equation}

\vspace{.2in}

 The convergents play important roles in the theory
of continued fractions. For instance, there exists the theorem of
inequalities,
\begin{equation}\label{33}
    n^{(1)}< n^{(3)}< \cdots  < n < \cdots n^{(4)} < n^{(2)}\, ,
\end{equation}
which provides a convenient algorithm to  address the question of
how best to approximate the continued fraction. In particular, we
observe that the exact distribution function of the B-anyons is
bounded between the first two convergents, thus:
\begin{equation}\label{34}
n^{(1)} < n < n^{(2)}\, .
\end{equation}
If we choose a value $q=\half$ for the purpose of illustration, we
obtain the result
\begin{equation}\label{35}
    \left ( \frac{3}{4 \ln 2} \right )\, \frac{1}{e^{\eta_i}-0.5} < n_i <
    \left ( \frac{3}{4 \ln 2} \right )\,\frac{1}{e^{\eta_i}-
    2.5}\,,
\end{equation}
valid when $q=\half$. It is important to stress that each of the
 quantities on the left, on the right
  and in the middle of the above inequality relation are exact. Moreover,
  Eqs.(\ref{31},\ref{32}) are exact results for arbitrary values of $q$.
   This is
 undoubtedly a remarkable result defined by the exact upper and lower
bounds. Moreover, the form of the distribution function is, other
than the numerical values, quite similar to the form used in the
ansatz of\cite{RAPNS1}.

\section{F-anyons}

In order to describe the Fermion-like anyons, we shall now
investigate the algebra defined by
\begin{equation}\label{36}
    a \adag + q^{-1} \adag a = q^{-N}\, ,
\end{equation}
together with the relations
\begin{equation}\label{37}
    [N,a]=-a, \quad [N, \adag ]= \adag \, ,
\end{equation}
where $N$ is the number operator, $0 \leq q \leq 1$ and the Fermi
limit is defined by $q \rightarrow 1$. This corresponds to the
case when we set $\kappa = -1$ in Eq.(\ref{2}). This algebra has
been investigated in the literature \cite{ALPNS2} but here we use
this algebra to describe interpolating statistics of Fermion-like
particles (F-anyons) with $q$ as the statistics determining
parameter.  Let us introduce the operator $\adag a = \hat{N}$ and
assume that the action on the Fock state can be described by
$\hat{N} |n \rangle = \beta_n |n \rangle$ where the eigenvalue
depends on $n$. We can determine $\beta_n$ as before. First we
observe that the relation $\hat{N} \adag + q^{-1}\adag \hat{N}=
\adag q^{-N}$ is true. We also know that the annihilation and
creation operators $a, \adag$ upon acting on state $|n \rangle$
lower and raise the number of quanta, with appropriate constants
to  be determined. As a consequence of this relation, we obtain
the result
\begin{equation}\label{38}
    \beta_{n+1} = q^{-n} - q^{-1} \beta_n \, .
\end{equation}
We can choose $\beta_0 =0$, thus defining the ground state as
vacuum and may then determine $\beta_n$ by solving the above
recurrence relation. We accordingly obtain
\begin{equation}\label{39}
    \beta_n = 0,\,  1 ,\, 0 , \,q^{-2}, \,0 ,\, q^{-4}, \cdots \, .
\end{equation}
Thus
\begin{equation}\label{40}
    \beta_n = \frac{1 - (-1)^n}{2}\, q^{-n+1} \, ,
\end{equation}
which reduces to $0$ and $q^{-n+1}$ respectively when $n$ is an
even or odd number. This behavior is totally in contrast with the
case of B-anyons.

\vspace{.2in}

From the action of the creation and annihilation operators on the
Fock states, we further obtain the results
\begin{equation}\label{41}
    \adag |0 \rangle = \sqrt{\beta_1}|1 \rangle = |1 \rangle; \, \quad
\adag \adag |0 \rangle = \sqrt{\beta_1} \sqrt{\beta_2}|2 \rangle
=0 \, .
\end{equation}
Consequently the sequence of states terminates and hence the Fock
states are $|0 \rangle, \, |1 \rangle $ only. Therefore in spite
of the interpolating statistics, the F-anyons obey Pauli exclusion
principle, just as ordinary Fermions do. This fact makes the
interpolating statistics in this investigation very different from
other theories.

\vspace{.2in}

Furthermore we observe that the algebra in Eq.(\ref{36}) does not
admit of a basic number as a solution of Eq.(\ref{38}), i.e.,
there is no basic number corresponding to the algebra in
Eq.(\ref{36}). The result $\adag a = [N], a \adag = [1-N]$ is not
true in the present case when the creation and annihilation
operators satisfy the algebra, Eq.(\ref{36}) in contrast to the
case in \cite{ALPNS2}. Instead, our analysis reveals the operator
relations
\begin{equation}\label{42}
    \adag a = \hat{N}= \frac{1 - (-1)^N}{2}\, q^{-N+1}
    , \quad a \adag = q^{-N} - q^{-1}\hat{N}\,
    ,
\end{equation}
Moreover, since the algebra is not related to basic numbers, the
theory of F-anyons does not require the use of JD and accordingly
we would employ the ordinary derivatives of thermodynamics.

\vspace{.2in}

From the definition of the mean value,
\begin{equation}\label{43}
    \hat{n}= \frac{1}{\cal Z} {\rm Tr} (e^{-\beta H} \adag a)\, ,
\end{equation}
proceeding as before, we obtain the result
\begin{equation}\label{44}
    \half (1- (-1)^n)= \frac{q^{-1}}{e^{\beta (E-\mu)} + q^{-1}}\,
    .
\end{equation}
We may re-express this to obtain the distribution function in the
form
\begin{equation}\label{45}
    n_i = \frac{2}{\pi}\arcsin \left ( \sqrt{\frac{q^{-1}}{e^{\eta_i} + q^{-1}} }
    \right )\, ,
\end{equation}
where $\eta_i = \beta (E_i-\mu)$. It is possible to express the
distribution function in the form of a series containing powers of
$g$, where
\begin{equation}\label{46}
    g= \frac{q^{-1}}{e^{\eta}+ q^{-1}}\, ,
\end{equation}
is a familiar form which occurs in the theory of
anyons\cite{RAPNS1} and hence this form is useful. Thus, using
\emph{Mathematica}, we obtain the result as a power series,
\begin{equation}\label{47}
    n= \frac{1}{\sqrt{g}} + \frac{7 \sqrt{g}}{6} + \frac{149 g^{3/2}}{120}
    + \frac{2161 g^{5/2}}{1680} + \cdots \, .
\end{equation}
This is the form from which we may obtain all the thermodynamic
functions for the F-anyons. For instance, the logarithm of the
partition function can also be expressed in the form of a power
series. However, a simpler approach prevails as we shall see
below.

\section{Thermostatistics of F-anyons}

The consequences of Eq.(\ref{47}) may be  rather intractable.
However, we can take advantage of the considerable simplification
that exists in the case of the F-anyons as follows, and thus there
is no need to deal with the series form. We recall that the Fock
states reduce to $n=0,1$ only. In this case, referring to
Eqs.(\ref{44},\ref{45}), $\sin^2 n \pi/2$ takes the values 0,1
only and hence can be replaced by n. Thus the distribution
function of Eq.(\ref{44}) simplifies to the form
\begin{equation}\label{48}
n_i= \frac{q^{-1}}{e^{\beta(E_i-\mu)}+ q^{-1}}\, .
\end{equation}
Other than the factor in the numerator,  we recognize  this to be
of the same form as an ansatz used in a previous
investigation\cite{RAPNS1}. It has the correct Fermi limit. We
shall now use this to investigate the thermostatistics. It is
important to stress here that in the case of F-anyons, we should
not use the JD but must employ ordinary derivatives.

\vspace{.2in}

We begin by observing that the logarithm of the partition function
is given by
\begin{equation}\label{49}
    \ln {\cal Z}= \sum_i \, \ln (1 + q^{-1}z e^{-\beta E_i})\, ,
\end{equation}
considerably different from the expression in Eq.(\ref{17}). We
also note that  partition function is explicitly dependent on the
parameter $q$.  Since
\begin{equation}\label{50}
    n_i = z \frac{\partial}{\partial z}\, \ln {\cal Z}\, ,
\end{equation}
this produces the distribution function given by Eq.(\ref{48}).
Upon converting the sum over states by an integration in the
standard manner, and introducing the thermal wavelength, we obtain
for the thermodynamic potential the expression
\begin{equation}\label{51}
    \Omega = -\frac{1}{\beta}\, \ln {\cal Z} = -\frac{1}{\beta
    \lambda^3}\, \ln (1 + q^{-1}z)
    - \frac{1}{\beta \lambda^3}
    \, f_{5/2}(q,z)\, ,
\end{equation}
where the function $f_n$  defined by
\begin{equation}\label{52}
    f_n(q^{-1}z)= \sum_{r=1}^{\infty}\, (-1)^{r+1}\,
    \frac{(q^{-1}z)^r}{r^n}\,
    ,
\end{equation}
is the generalization of the Riemann Zeta function for the
Fermion-like particles. All of these functions reduce to the
standard Fermion case in the limit when $q \rightarrow 1$. The
first term corresponds to the fact that we have isolated the zero
momentum state in the standard manner. The pressure is obtained in
the thermodynamic limit as
\begin{equation}\label{53}
    P = \lim_{V \rightarrow \infty, N \rightarrow \infty}
     \left ( - \frac{\Omega}{V} \right )\, .
\end{equation}
We thus determine the pressure of the F-anyons to be
\begin{equation}\label{54}
    P= \frac{1}{\beta \lambda^3}\, f_{5/2}(q^{-1}z)\,
\end{equation}
which agrees with the standard expression\cite{Reichl} in the
Fermi limit. In a similar manner we determine the mean density to
be given in the thermodynamic limit by
\begin{equation}\label{55}
\frac{n}{V}= \frac{1}{\lambda^3}\, f_{3/2}(q^{-1}z)\, .
\end{equation}

\vspace{.2in}

We can learn how these thermodynamic quantities for F-anyons
compare with the corresponding ones for ordinary fermions by
studying the graphs \cite{Reichl} for the functions $f_{3/2},
f_{5/2}$. Observing that $f_n(q^{-1}z) > f_n(z)$, it thus follows
for instance that the pressure of the F-anyon gas is greater than
that of ordinary fermions at the same temperature and for the same
fugacity. We observe that this conclusion is qualitatively the
same as the conclusions for the q-fermions discussed
in\cite{ALPNS2}. However, it must be stressed that $q$ in
\cite{ALPNS2} refers to arbitrary values  whereas here the
parameter $q$ is restricted to be in the range $0 \leq q \leq 1 $.
 The actual expressions are different and  the results
follow from a different algebra.

\vspace{.2in}

Next we may investigate the virial expansion. In the standard
notation, we obtain the result
\begin{equation}\label{56}
    \frac{P v}{k T}= 1 + \frac{1}{2^{5/2}} \left ( \frac{\lambda^3} {v} \right )
    + \left ( \frac{1}{8}- \frac{2}{3^{5/2}} \right ) \left
    ( \frac{\lambda^3} {v}\right )^2 +
    \cdots \, .
\end{equation}
It is interesting to note that the virial coefficients are
independent of $q$. Indeed it is the same as for ordinary
fermions\cite{Reichl}. This situation contrasts with the
conclusions of\cite{RAPNS1} where the ansatz for the distribution
function was introduced differently and where the calculations
were done in two dimensional space and it is also different from
the conclusion in \cite{ALPNS2}.

\vspace{.2in}

We can obtain the internal energy in the form
\begin{equation}\label{57}
    U= \frac{3 k T V}{2 \lambda^3}f_{5/2}(q^{-1}z)\, ,
\end{equation}
which has the correct Fermi limit.  For the entropy of the
F-anyons we obtain the expression
\begin{equation}\label{58}
    \frac{S}{N k}= \frac{5}{2}\, \frac{f_{5/2}(q^{-1}z)}{f_{3/2}(q^{-1}z)}
    - \ln z \, ,
\end{equation}
which agrees with the result for standard fermions in the fermi
limit. We also observe that for $q \neq 1$ the entropy is larger
than that of  ordinary fermions.

\vspace{.2in}

We may conclude with some general results for the F-anyons. In the
limit of large energy, the F-anyon distribution function reduces
to
\begin{equation}\label{59}
    n_i \longrightarrow q^{-1} e^{- \beta E_i}\, ,
\end{equation}
which, other than the normalization factor, is the same as in
classical Boltzmann case. In the limit when $E=\mu$, the
distribution reduces to
\begin{equation}\label{60}
    n_i= \frac{q^{-1}}{1 + q^{-1}} \geq \half \, ,
\end{equation}
which takes the value $\half $ only in the Bose limit when $q=1$.
In the high temperature limit, when $T \rightarrow 0$, it is clear
that the distribution function reduces to the standard unmodified
step form for all values of $q$. Hence the interpolation
statistics may be interpreted as solely a finite temperature
effect. The modification at higher temperatures is similar to the
standard Fermions except that the parameter $q$ also plays a role.

\vspace{.2in}

The dependence on the parameter $q$ is somewhat subtle for many of
the thermodynamic functions. As an illustration, we may consider
the dependence of the Fermi-energy in some detail. The number
density of F-anyons is given by the distribution function
\begin{equation}\label{61}
    \frac{N}{V}= \frac{g}{\lambda^3}\, f_{3/2}(q^{-1}z)\, ,
\end{equation}
where $g$ is the multiplicity factor. This can be expressed by the
series
\begin{equation}\label{62}
\frac{N}{V}= \frac{4 \pi g}{3}(\frac{2m k T}{h^2})^{3/2} (\ln
(q^{-1}z)^{3/2})\left ( 1 + \frac{\pi^2}{8} (\ln(q^{-1}z)^{-2})+
\cdots \right )\, .
\end{equation}
This can be employed to determine the chemical potential $\mu$ of
F-anyons in terms of the Fermi-energy of standard Fermions,
\begin{equation}\label{63}
    E_F=\frac{3 N}{4 \pi g V}^{2/3}\, \frac{h^2}{2m}\, .
\end{equation}
In the lowest approximation, we obtain
\begin{equation}\label{64}
    \mu = \frac{3 N}{4 \pi g V}^{2/3}\, \frac{h^2}{2m}- k T \ln
    {q^{-1}}\, ,
\end{equation}
which shows that the $q$-dependence appears only at finite
temperatures. The expression beyond the zeroth approximation is
given by
\begin{equation}\label{65}
\mu = - k T \ln {q^{-1}} + E_F \left ( 1 - \frac{\pi^2}{12}\,
\left ( \frac{kT}{E_F} \right )^2 + \cdots
 \right )\, .
\end{equation}
Thus the temperature dependence of the chemical potential of the
F-anyons is different from that of standard Fermions due to
dependence on the statistics determining factor, and it is
relatively decreased.

\section{Concluding remarks}

In summary, we have addressed the following problem: what kind of
oscillator algebra can adequately describe the statistics which
interpolates continuously between Bose and Fermi statistics? This
is of considerable interest due to the fact that any connection
between particles obeying fractional statistics, anyons, and
deformed Lie algebras (quantum groups) has been looked at in the
literature with disfavor, other than in two space dimensions.
Indeed, many investigations described the anyons by the braid
group because of the connection between permutation symmetry and
the rotation group in two space dimensions. We have thus
established that interpolating statistics  existing in ordinary
3+1 dimensions can arise from deformed quantum oscillator algebra.

\vspace{.2in}

Starting from the quantum group corresponding to $q$-deformed
oscillators, described by one algebra containing two parameters
$q$ and $\kappa$, we interpret the parameter $q$  as the
statistics determining parameter, which then delineates the
algebra into two distinct regimes for $\kappa=1$ and $ \kappa=-1$
corresponding to Boson-like and Fermion-like anyons. We have
investigated the two distinct algebras in detail and derived the
consequences.

\vspace{.2in}

We have investigated the thermostatistics of the B-anyons in order
to ascertain how they differ from ordinary Bosons. Various
thermodynamic functions have been determined. We have been able to
express the distribution function in this case in terms of
continued fractions. One advantage of the continued fraction form
is that we can obtain convenient upper and lower bounds for the
distribution function. As an example, we were able to obtain exact
upper and lower bounds of the distribution function of the
B-anyons for arbitrary values of $q$. We have obtained the virial
expansion for B-anyons. We also obtain the second virial
coefficient to establish its dependence on the statistics
parameter via basic numbers.

\vspace{.2in}

We have examined in detail the algebra which gives rise to the
interpolating statistics for the F-anyons. While the B-anyon
algebra leads to basic numbers and the $q$-calculus based on
Jackson derivatives, the features of the algebra of F-anyons are
quite different: it does not lead to basic numbers; rules of
ordinary derivatives prevail; there are only two Fock states, and
hence F-anyons obey Pauli exclusion principle just as ordinary
Fermions do. We have studied the thermostatistics of F-anyons as a
consequence of this algebra. The thermodynamic properties of
F-anyons are quite different from q-Fermions, although both of
these depend on the parameter. For instance we are able to show
that the virial coefficients of the F-anyons are no different from
those of the standard Fermions, while that is not the case for
$q$-deformed Fermions. We have obtained the temperature dependence
of the Fermi energy of F-anyons in the form of a power series and
show that the Fermi energy depends linearly on temperature in the
first approximation which vanishes in Fermi limit, $q \rightarrow
1$.

\vspace{.2in}

 Some investigations in the past \cite{Greenberg}
considered interpolation in order to accommodate a deviation from
Pauli exclusion principle and tended to develop a theory to
account for any experimental evidence along these lines. Here in
the present investigation we have discussed different situations
including the consequence that Pauli principle is strictly valid
as for ordinary Fermions. This might indicate, that contrary to
appearances, our algebra is different from the original idea of
Greenberg and others even at the domain of the single level
system. It is rather surprising that the consequences of a similar
algebra for B-anyons and F-anyons would be so strikingly
different. It is interesting that in this theory, Pauli exclusion
principle is valid for F-anyons while it has to be imposed by hand
explicitly or implicitly for $q$-deformed Fermions \cite{ALPNS2}.
It is also interesting that F-anyons would have a two dimensional
Fock space, that Pauli principle would be valid for a range (dense
set) of the statistics parameter.

\vspace{.2in}

We believe our investigation provides a confirmation of the notion
that there is a close connection between quantum Lie algebra of
deformed quantum harmonic oscillators and intermediate statistics
as well as the concept that interpolation statistics is not
related to 2+1 dimensions of space-time.

\acknowledgments

I would like to express my gratitude to A. Lavagno of Politecnico
di Torino, Torino, Italy, for fruitful discussions and
correspondence on the subject of interpolating statistics and
q-deformed oscillator algebra.

\end{document}